# Near-field radiative heat transfer between hybrid polaritonic structures

Naeem Iqbal[1,+], Sen Zhang[1,+], Shuai Wang[2,+], Zezheng Fang[3], Yaoyuan Hu[2], Yongdi Dang[1], Minjie Zhang[3], Yi Jin[1], Jianbin Xu[4], Binfeng Ju[2,*] and Yungui Ma[1,*]

[1]*State Key Lab of Modern Optical Instrumentation, Centre for Optical and Electromagnetic Research, College of Optical Science and Engineering, International Research Center for Advanced Photonics, Zhejiang University, China*
[2]*The State Key Lab of Fluid Power Transmission and Control, School of Mechanical Engineering, Zhejiang University, Hangzhou 310027, China*
[3]*Zhejiang Province Key Laboratory of Quantum Technology and Device, Department of Physics, Zhejiang University, Hangzhou, People's Republic of China*
[4]*Department of Electrical and Electronic Engineering, The University of Hong Kong, Pokfulam Road, Hong Kong, China*

[+]*These authors contributed equally to this work.*
*\*E-mail address: yungui@zju.edu.cn (Y. Ma); mbfju@zju.edu.cn*

**Abstract**

Near-field radiative heat transfer between close objects may exceed the far-field blackbody radiation in orders of magnitude when exploiting polaritonic materials. Great efforts have been made to experimentally measure this fundamental stochastic effect but mostly based on simple materials. In this work, we foster an all-optical method to characterize the heat transfer between less explored plasmon-phonon hybrid polaritonic systems made of graphene-SiC heterostructures. A large heat flux about 26 times of the blackbody radiation limit is obtained over a 150-nm vacuum gap, attributed to the couplings of three different surface modes (plasmon, phonon polaritons and frustrated mode). The interaction of polaritonic modes in the hybrid system is also explored to build a switchable thermophotonic device with nearly unity heat flux tunability. This work paves the way for understanding mode-mediated near-field heat transfer and provides a platform for building thermophotonic or thermo-optoelectronic blocks for various applications.

**Introduction**

Near-field radiative heat transfer (NFRHT) between closely spaced objects may exceed the far-field Planck's blackbody radiation limit in orders of magnitude at the contribution of evanescent waves [1, 2], in particular with bounded modes originated from surface plasmon polaritons (SPPs) [3, 4], surface phononic polaritons (SPhPs) [5, 6] or hyperbolic materials [7-9]. They may find essential applications in thermal rectification [10-13], imaging [14, 15], radiative cooling [16-18] or thermophotovoltaics [19-22]. In recent years, new ideas enabled by two-dimensional (2D) materials or artificial nanostructures with exotic infrared features have been vastly explored to deepen the understanding on near-field thermophotonics [23-27]. Among them, the monolayer graphene has received special attentions in tailoring thermal photons due to their outstanding features like strong SPPs, tunable conductivity, high integration freedom with other species, etc. [28, 29]. For example, it could induce hybridized polaritons to gain versatile physical features when coupled with a polar material [30-32] or provide a heterojunction to collect thermal photons when integrated with a semiconductor [33, 34]. However, on the other side, it is still a rather challenging task to precisely characterize these exotic effects with high reliability, although we have witnessed a large advancement for the measurement skills of NFRHT in the last decade [35-39]. The gap distance should be small enough to allow for the tunneling of highly localized SPPs in between graphene. Limited by the substrate flatness, the sample size often has to be controlled to facilitate the alignment. In previous measurement setups [9, 35-37, 40-45], electric contacts have been largely used to control temperatures and sense heat flux. This may be challenged in many scenarios for small samples where electric contacts may cause heat loss conduction with amplitude equivalent to or even stronger than the radiation channel. This will degrade the measurement accuracy in counting the net radiative heat flux. New measurement architecture is basically needed, for 2D materials or metamaterials which are usually available only in millimeter or even smaller sizes.

In this work, we developed an all-optical approach to characterize the near-field thermal radiation particularly between plasmon-phonon hybrid photonic nanotructures made of graphene (gra) covered polar semiconductor 6H-SiC, i.e., forming gra-SiC heterostructure via Van der Waals force. The working area is $1\times1$ mm$^2$ with a gap distance of 150 nm helped by nanopillar spacers. This geometry is very suitable for graphene or metamaterials fabricated by advanced techniques like electron beam lithography (EBL). We obtained a large measured heat flux exceeding the blackbody radiation limit by a factor of 26 folds at a temperature gradient of 45K, in good agreement with the prediction of the thermal electrodynamic theory. The contribution from three conduction channels arising from the couplings of SPP, SPhP and frustrated modes was explicitly identified from the spectra of local density of states (LDOS) and transmission coefficients (TCs). The interaction of SPP and SPhP modes in the hybrid system was also utilized to develop a switchable thermophotonic system with nearly unity heat flux tunability.

**Experimental procedure and characterization**

As mentioned above, NFRHT between 2D materials has been extensively investigated in theories. Our current intention is mainly on the experiment part, which is less studied especially for hybrid polaritonic structures. We explore the heterostructure made of graphene and polar semiconductor silicon carbide (SiC), which is of strong SPP and SPhP responses in long-infrared spectrum. Schottky junction may also be expected at the interface due to their relative work functions [46, 47], which may offer an unprecedented measure for future to control and recycle the near-field heat flux. In this work, intrinsic SiC single crystal substrates are utilized for the purpose to understand the plasmon-phonon interaction and their influence on thermophotonic radiation. Figure 1a gives a schematic for the all-optical wireless near-field heat flux measurement setup developed here for small samples. The emitter (top) has a total area of $1 \times 1$ mm$^2$ far smaller than the receiver (bottom) ($16 \times 16$ mm$^2$). These two ends are separated at a fixed gap distance of 150 nm with the help of photoresist (SU-8, thermal conductivity of 0.3 Wm$^{-1}$K$^{-1}$) nanopillars [48]. Similarly, as before, magnets in both the emitter and receiver sides have been added to apply a proper contacting force[9, 33]. A commercial graphene monolayer grown by chemical vapor phase deposition (CVD) (7440-44-0, XFNANO Mater Tech Co., Ltd, China) is utilized here by following the standard transfer procedure [49]. In our device, the emitter is heated by an external laser focused onto one side of magnet ($1\times1\times1$ mm$^3$) coated with the blackbody paste and the receiver is maintained at room temperature utilizing a thermoelectric (TE) cooler (1- 12705, Realplay, China). The temperatures of emitter and receiver are in-time monitored by a top thermograph (HPM 11, HIKIMICRO, China) (Figure 1b) via a top infrared window at an uncertainty of 0.1 K. An infrared microdomain lens has been applied here to acquire a high spatial resolution ~ 50 μm. The whole experiment is performed in a 1000 class clean room while the NFRHT measurement is conducted at a vacuum pressure of $\sim 10^{-5}$ Pa. As indicated by the equivalent thermal circuit (Figure 1c), the loaded laser input power $P_{in}$ is dissipated through two paths: background loss $P_b$ and thermal radiation $P_{rad}$, i.e., $P_{in} = P_{rad} + P_b$. The background loss includes the heat conduction along the supporting nanopillars $P_{AZ}$ and the far-field radiation into the surrounding environment $P_{ff}$. The latter is independent of gap distance [43] and evaluated using the Wien distribution law. By our setup, the near-field heat flux is obtained via $P_{rad} = P_{in} - P_{ff} - P_{AZ}$.

In our measurement, the maximum temperature variation is controlled to be smaller than 50K. Within this range, the variation of infrared properties of both 6H-SiC substrate and graphene are neglected. The surface curvature including the roughness of gra-SiC sample is smaller than 14 nm as measured by a 3D optical profiler (NewView 8000 Series, ZYGO, America) over the area of $1 \times 1$ mm$^2$ (Figure 2a). The Raman spectrum of the sample is given in Figure 2b. Several peaks between 1400 and 1800 cm$^{-1}$ are observed arising from the longitudinal and transverse optical phonons of 6H-SiC (0001) (lattice texture is decided by X-ray diffraction measurement). For graphene, typical G-band peak from the in-plane lattice vibration is observed at 1591 cm$^{-1}$ and the double-resonant 2D-band peak at 2697 cm$^{-1}$, consistent with earlier values reported for epitaxially-grown monolayer graphene on SiC substrate [50]. The monolayer nature of graphene can be derived from the strong 2D peak [51]. In addition, the absence of D-band peak implies our graphene monolayer has very weak residual defects. The Fermi level ($E_F$) estimated from the G-mode frequency is about 0.26 eV, which is typical for a CVD grown graphene [52].

The reflection spectra of the samples measured Fourier transform infrared spectroscopy (FTIR) are plotted in Figure 2c. For better visualization, the reflectivity of the gra-SiC sample is shifted upward by 0.05. As indicated by the dashed lines, Lorentz model for SiC ($\varepsilon_{SiC} = \varepsilon_\infty \left[\frac{\omega^2 - \omega_{LO}^2 + i\omega\gamma}{\omega^2 - \omega_{TO}^2 + i\omega\gamma}\right]$) and Drude model for graphene ($\sigma_{gra} = \frac{ie^2|E_f|}{\pi\hbar^2(\omega + i\tau^{-1})}$) have been utilized to fit the measured reflection data. From this operation, we get the parameters $\varepsilon_\infty = 6.7$, $\omega_{LO} = 1.8355 \times 10^{14}$ rad/s, $\omega_{TO} = 1.4989 \times 10^{14}$ rad/s, $\gamma = 8.972 \times 10^{11}$ rad/s for SiC and in Figure 2d, plot the permittivity spectra exhibiting a prominent acoustic resonance around 797 cm$^{-1}$. For graphene, we have $E_F = 0.26$ eV and $\tau = 100$ fs. The fitted Fermi level is in good agreement with that derived from the Raman spectral peak. Figure 2e plots the real and imaginary conductivities of graphene as a function of wave number. A peak around 52.2 cm$^{-1}$ is found for Re($\sigma_{gra}$), which is much lower than the acoustic resonance of SiC in frequency. Note the high dielectric constant of SiC has red shifted the plasmonic band of graphene compared with vacuum background. Inversely, as seen later, the plasmonic response of the graphene monolayer will also affect the phononic dispersion of SiC, thus realizing a hybridization nature. As the key parameters of near-field radiation, the LDOS of the heterostructure is mutually decided by the plasmonic and phononic responses of the constituent materials.

**Results and Discussion**

The radiative flux density (*h*) between SiC substrates with and without graphene cover is measured by exploiting our newly designed laser based all-optical setup for several temperature differences (Δ*T*) across the vacuum gap. The measured results are shown in Figure 3, where we see the introduction of graphene could greatly raise the slope of the *h*~Δ*T* curve. Here, the temperature of receiver is fixed at 300 K. A gap distance of 150 nm between the emitter and receiver is attained through nanopillars height with uncertainty of 11 nm. Such uncertainty in gap estimation measurement is represented by bandwidth of color band in theoretical calculation. Also, the temperature measurement uncertainty is extrapolated about 1 K. The conduction through pillars remained smaller than one third of measured radiative power. For comparison, theoretical results using fluctuation electrodynamic and blackbodies radiative flux are computed. A good agreement is found between the measured and computed results for samples with and without graphene, which show prominent super-Planckian thermal radiation effect. For example, as indicated by the arrow, at a temperature difference of 45 K, it was found that radiative flux density between SiC covered with graphene is enhanced by a factor of ∼26 folds than the flux density between the blackbodies intimating the involvement of thermally agitated evanescent waves. While in case of SiC without graphene this enhancement factor is appraised to be of ∼12 fold for a temperature gradient of 40 K. Indeed, by adding the graphene, its role in near-field enhancement as a result of coupling of SPPs is vital as evidenced here.

The preliminary physics explaining the near-field enhancement can be understood from the investigation of the local density of state and energy flux spectra. In this context, local density of states is determined at a distance of 150 nm in free-space from the interface of SiC with and without graphene as shown in Figure 4a. Here onward temperature difference and vacuum gap separation are fixed at 45 K and 150 nm, respectively. The resonance peak from the surface phonon polaritons appears at $\omega_{\text{SPhPs}} \approx 951\,\text{cm}^{-1}$ (At this point, $\varepsilon_{\text{SiC}}$ = -1.04+$i$·0.12) which is entirely related to the electric field energy contribution [53]. The density of state described by energy density of blackbody and spontaneous emission rate can be strongly modulated by surface modes [54, 55]. Introduction of graphene will provide a strong measure to modulate the surface optical energy density and thus the near-field heat flux using its chemical potential tunability. As shown in Fig. 4a, the acoustic resonance peak of SiC is broadened and blue shifted to a higher frequency $\sim$968 cm$^{-1}$ when covered by the monolayer graphene, which has a metallic response in this frequency band. A second DOS peak around $\omega \approx 792\,\text{cm}^{-1} = \omega_{\text{T}}$ is also observed, which is independent of the existence of graphene [56]. At the low frequencies around 255 cm$^{-1}$, a broad DOS peak is observed owning to the SPP resonance of graphene. Note the plasmonic mode band of graphene is strongly modulated by the high-index SiC substrate. Therefore, the mutual interaction to each other's dispersion in the plasmon-phonon hybrid system is evidenced, which as discussed later, will give rise to a Fermi level and gap distance dependent near-field heat flux.

The heterostructure employed here could support three modes inside the materials: propagating modes ($k_{\parallel} < k_0$), frustrated modes ($k_0 < k_{\parallel} < \varepsilon k_0$) and surface polaritonic modes ($k_{\parallel} > \varepsilon k_0$) with quantities $\varepsilon$, $k_{\parallel}$ and $k_0$ denoting material dielectric constant, in-plane wave vector and free-space wave vector. Frustrated modes are evanescent in vacuum and propagating inside the material while surface modes are evanescent both inside vacuum and material in which surface modes give major contribution to radiative flux. Note that in Figures 4b and 4c, we use band 1, band 2 and band 3 to denote the frequency ranges $(0 - 400)$ cm$^{-1}$, $(600 - 800)$ cm$^{-1}$ and $(900 - 1100)$ cm$^{-1}$, which correspond to the band of SPPs of graphene, frustrated modes of bulk SiC and SPhP of SiC, respectively. The near-field field coupling between surface modes abetted by emitter and receiver leads to symmetric and antisymmetric modes distribution through dispersion relation as apparent from the characteristics of transmission coefficients (TCs) shown in Figure 4b (SiC without graphene) and 4c (SiC with graphene). For larger $k_{\parallel}$, these splitting of modes in vacuum from both configurations merge into a single mode at respective resonance frequency due to the weakened inter-surface coupling. The wide spread 2$^{\text{nd}}$ band representing the losses where value of imaginary part of dielectric constant becomes huge $\sim$552. On the other hand, in the first and third band such losses are small. Moreover, in the first band of TCs, only graphene's low energy SPPs contribute to coupling mechanism with smaller cut-off wave vector as compared to the third band (SPhP mode has a lower loss). By comparing the top panels in Figures 4b and 4c, we see that the existence of monolayer graphene will blue shift the SiC's super-modes and thus modulate the heat flux.

The radiative flux as a function of frequency is obtained by integrating TCs over the whole in-plane wave vector and the result is depicted in Figure 4d. A narrow emission peak in close proximity to the phononic resonance is observed using bare SiC substrates; the inclusion of graphene makes the difference in radiative flux mediated by surface plasmons polaritons as apparent from the inset figure for flux density obtained by integrating radiative flux over three different frequency bands. The excitation of SPPs due to the presence of graphene over SiC in the first band and their coupling in vacuum contributes to a larger amount of radiative flux as compared to the rest of the bands. Moreover, the third band phononic emission peak not only is blue shifted but becomes relatively wider in width as well due to the relatively larger loss level of graphene. In this context, it is worth mentioning that the graphene plasmons wavelength is an important parameter which is controlled by applied gate voltages which sets the length scale in its surrounding medium over which it can influence the electrodynamics of surrounding. Moreover, the plasmonic dispersion feature of graphene highly remains preserved outside the phononic band as apparent from band 1 of Figure 4c [57, 58].

**Conclusion**

Near-field teat transfer between gra-SiC heterostructures has been investigated in this work using a home-made all-optical measurement setup. The accuracy of the results is verified by their good agreement with the theoretical predictions. At a gap distance of $150 \pm 11$ nm and temperature gradient of 45 K, we obtained a high radiative flux surpassing the far-field blackbody radiation limit by 26 folds. The energy transmission spectral analysis indicates that there exists a strong interaction between the monolayer graphene cover and the SiC bulky substrate. The introduction of graphene could greatly enhance the heat flux by contributing a highly-localized low-frequency plasmonic coupling band. More layers consisted of (2D) materials with different band responses may be considered in the future to further enhance the heat conduction channels for stronger heat flux. It is also very interesting to check the tunability of the total heat flux via controlling the voltage bias of graphene in particular at smaller gap distance. Regarding the experimental technique, an infrared microscope and heat laser may be added before the thermal imager so that smaller pieces of samples with diverse of optical properties could be potentially investigated in the near-field thermophotonic regimes. Alternatively, in the future, temperature dependent Raman or photoluminescent spectral imaging techniques may be utilized but with high spatial resolutions. The current work has implication in providing a pronounced measure to exploit the compelling thermophotonic properties and device applications enabled by natural and artificial materials in infrared bands. Moreover, doped SiC substrate (polar semiconductor) with both plasmonic and phononic features could be utilized and its combination with graphene will also induce an interface Schottky junction rendering a thermo-optoelectrical transfer channel. This junction may be utilized to sense the transient near-field heat flux for example through the dark current measurement or harvest the near-field thermo-photon energy into electricity.

**Declaration of Competing Interest**

The authors declared there is no competing interest conflict.

**Acknowledgement**

The authors thank the partial supports from National Natural Science Foundation of China 62075196 and 61775195, Natural Science Foundation of Zhejiang Province Z21F050013, National Key Research and Development Program of China 2017YFA0205700 and Fundamental Research Funds for the Central University.

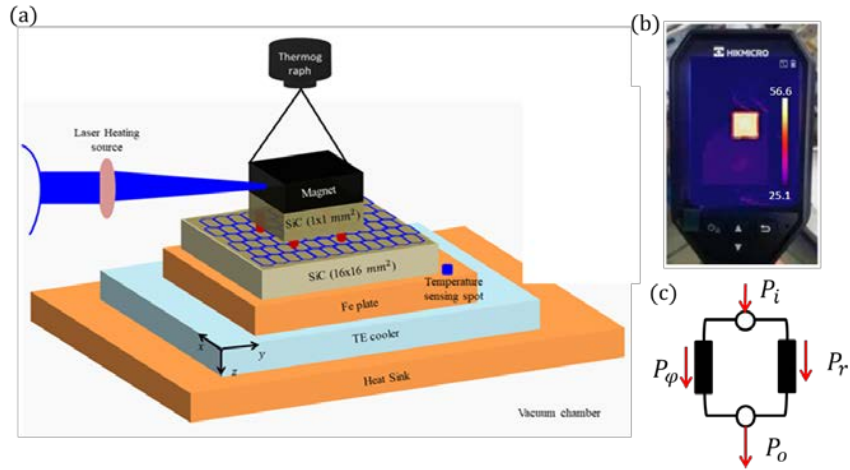

Figure 1: Measurement setup. (a) Schematic of the NFRHT measurement setup, where a laser heating source is focused on the emitter and the temperatures are captured by a thermograph. (b) The image for temperature measuring thermograph. (c) The equivalent measurement thermal circuit. The input laser power $P_{in}$ is divided into the near-field radiation $P_{rad}$ and the background $P_b$. The latter includes the far-field radiation power $P_{ff}$ and the heat conduction along the spacing photoresist pillars $P_{AZ}$. The near-field heat flux is calculated by $P_{rad} = P_{in} - P_{ff} - P_{AZ}$.

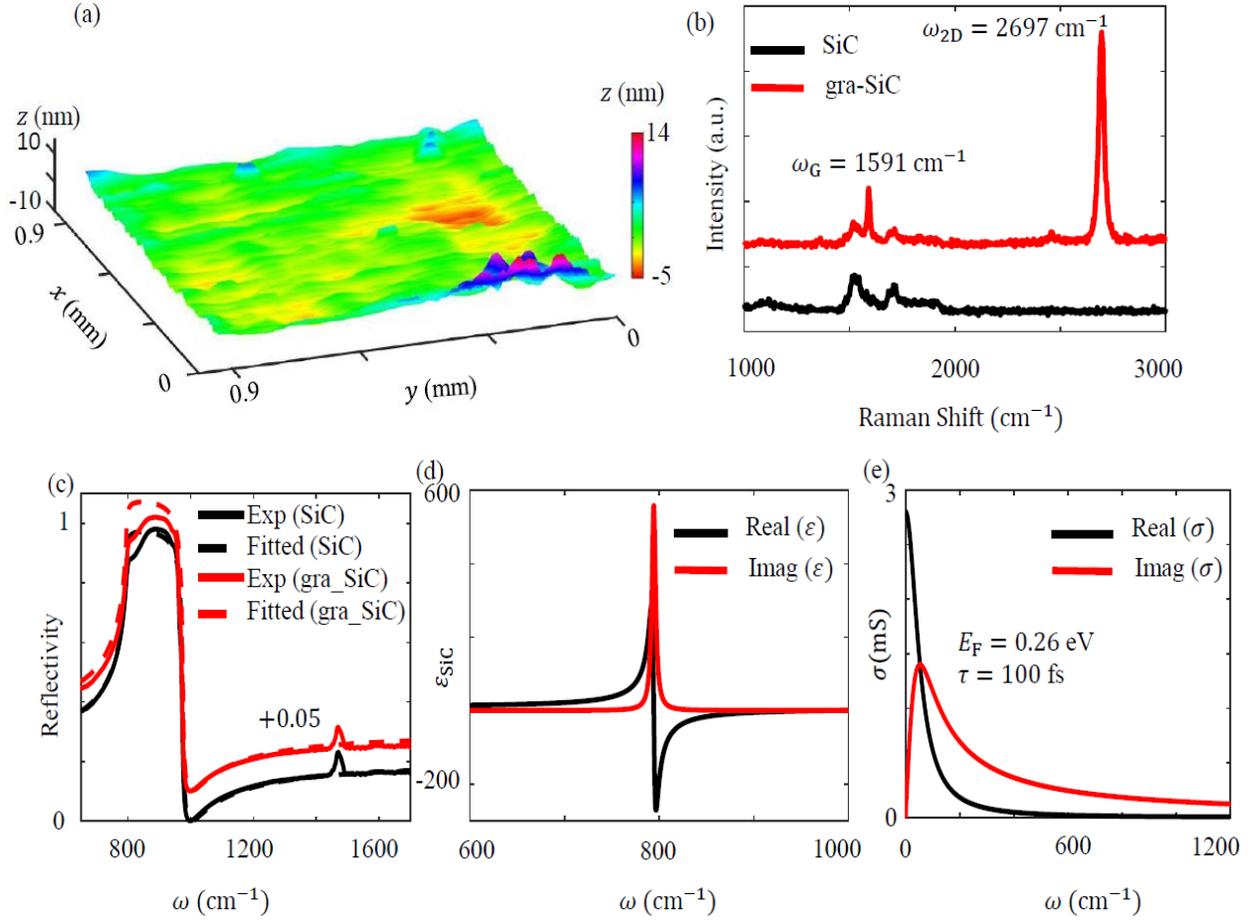

Figure 2: Material property characterization. (a) The roughness and surface curvature characterization of 1×1 mm² gra_SiC heterostructure. (b) Raman spectrum of SiC without (black) and with (read) graphene. For graphene, two peaks at $\omega_G = 1591$ cm⁻¹ and $\omega_G = 2697$ cm⁻¹ are indicated. (c) Measured reflectivity for SiC without (solid black) and with (solid red) graphene. Respective dashed lines are obtained using Lorentz dielectric constant model for SiC and Drude dielectric constant model for graphene. A multiplier by a factor of 0.05 for the heterostructure is used for better visualization. (d) Real and imaginary parts for dielectric constant derived for SiC (fitted parameters: $\varepsilon_\infty = 6.7$, $\omega_{LO} = 1.8355 \times 10^{14}$ rad/s, $\omega_{TO} = 1.4989 \times 10^{14}$ rad/s, $\gamma = 8.972 \times 10^{11}$ rad/s). (e) Real and imaginary parts of conductivity derived for graphene (fitted parameters: $E_F = 0.26$ eV and $\tau = 100$ fs).

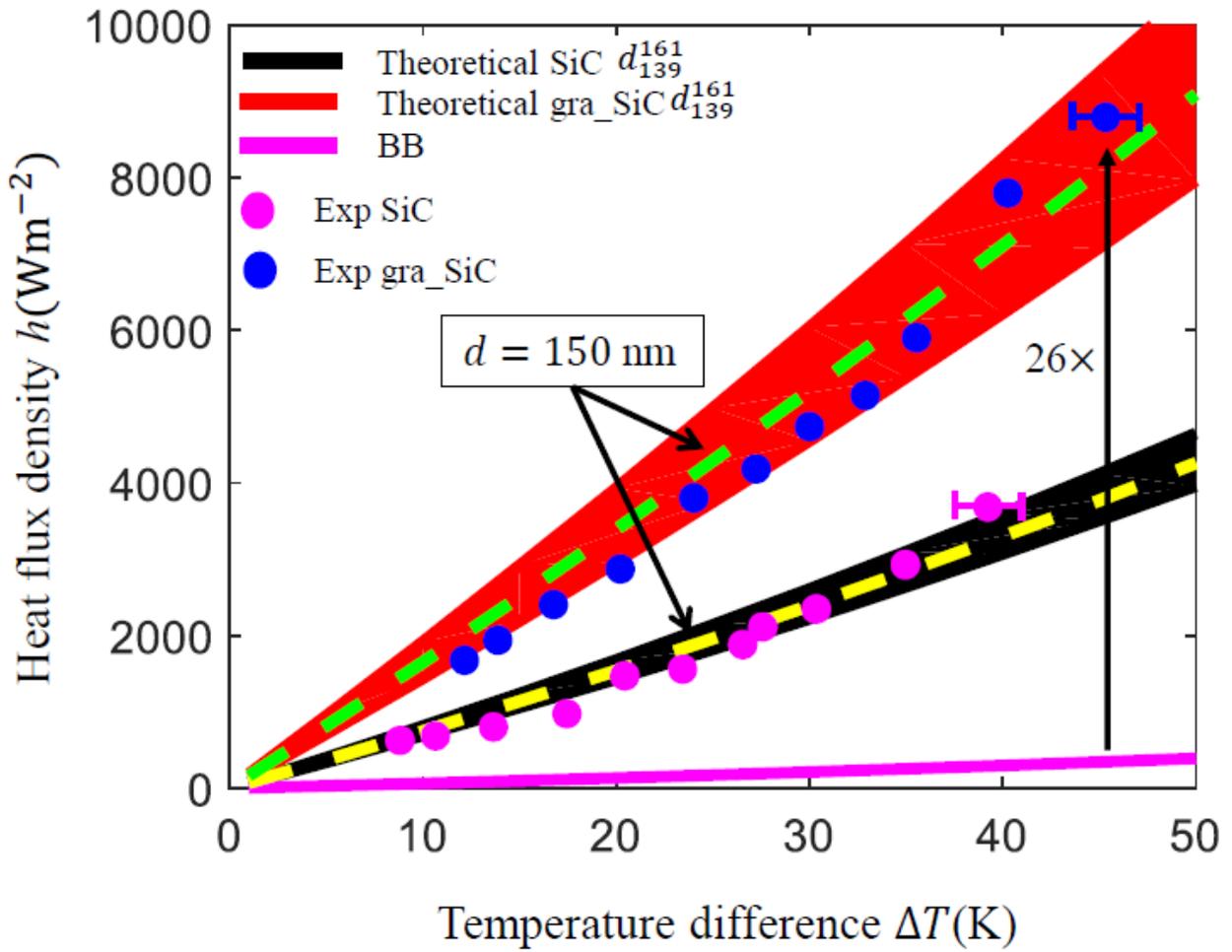

Figure 3: Measured and simulated results. (a) Theoretical and experimental NFRHT results between identical emitter and receiver made of SiC with (blue) and without (purple) graphene as a function of temperature gradient. The experimental results are indicated by circles with bars denoting the measuring uncertainty. The red and black shaded color regions are numerically calculated using the classic fluctuational electrodynamics theory incorporating the measurement uncertainties in material and structural parameters. The blackbody (BB) radiation (purple) is also given for comparison. The near-field flux density of the heterostructure exceeds the blackbody radiation limit by a factor of 26 at a bias of 45K.

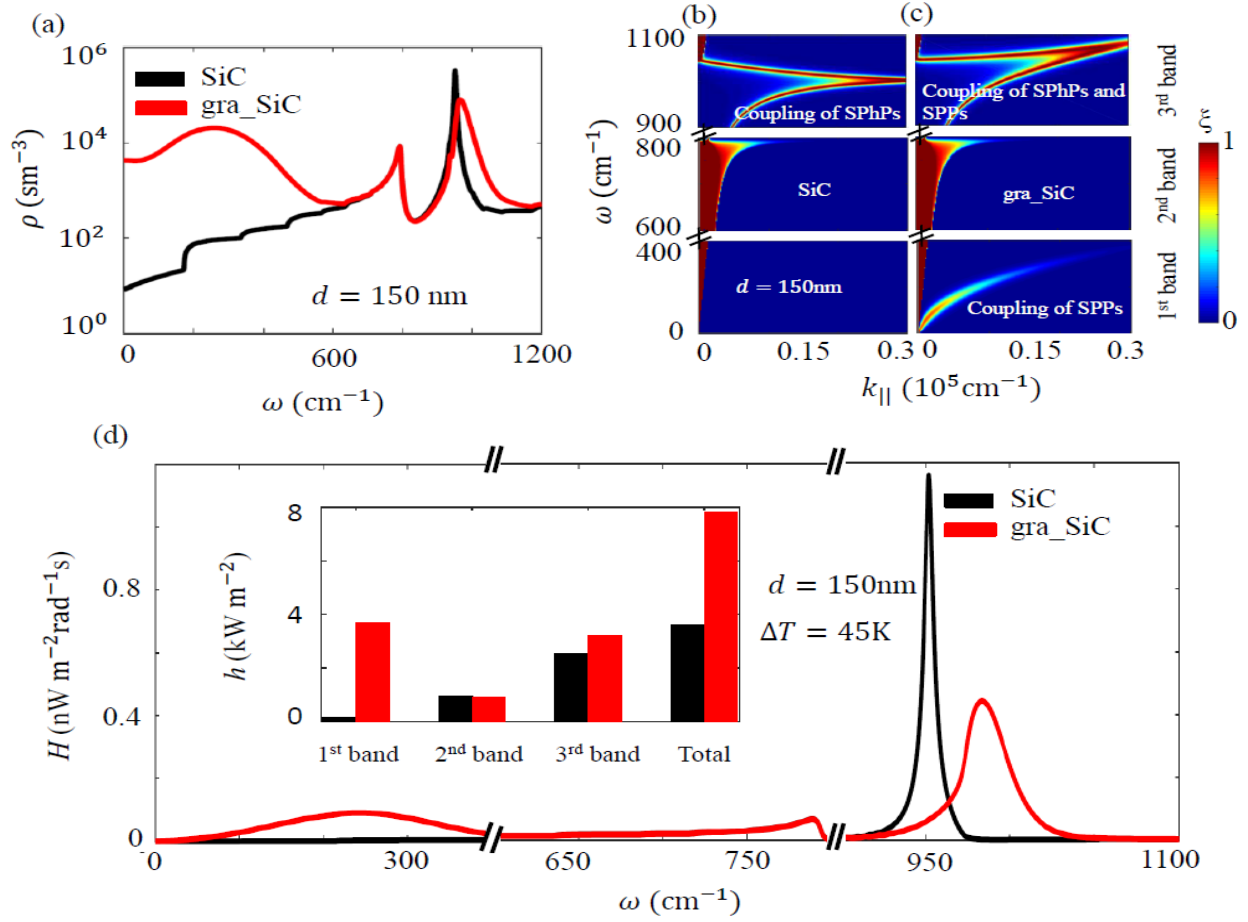

Figure 4: Spectral mode analysis. (a) LDOS calculated at a distance of 150nm in free-space from the interface of SiC with (red) and without (black) graphene. (b) and (c) The energy transmission coefficient $\xi$ for both configurations, i.e., SiC without and with graphene, respectively, as a function of in-plane wave vector $k_\parallel$ and angular frequency $\omega$. We divide the *y*-axis spectrum into three sub-bands according to their resonance natures. (d) Energy flux density spectra for a separation of 150 nm and temperature gradient of 45 k. The inset figure represents the flux density integrated for three different sub-bands, where the first band arising from the SPP mode of graphene gives larger contribution to radiative flux as compared to the rest bands.